\documentclass{article}
\usepackage[preprint]{unireps_2023}
\setcitestyle{numbers}
\setcitestyle{comma}
\setcitestyle{square}
\usepackage[utf8]{inputenc} % allow utf-8 input
\usepackage[T1]{fontenc}    % use 8-bit T1 fonts
\usepackage{hyperref}       % hyperlinks
\usepackage{url}            % simple URL typesetting
\usepackage{booktabs}       % professional-quality tables
\usepackage{amsfonts}       % blackboard math symbols
\usepackage{nicefrac}       % compact symbols for 1/2, etc.
\usepackage{microtype}      % microtypography
\usepackage{xcolor}         % colors
\usepackage[version=4]{mhchem}
\usepackage{graphicx}
\usepackage{dirtytalk}
\usepackage[skip=1.8pt]{caption}
\usepackage[T1]{fontenc}
\usepackage{xr}

\title{Unsupervised learning on spontaneous retinal activity leads to efficient neural representation geometry}

\author{Andrew Ligeralde$^{1, 2*}$ \quad Yilun Kuang$^{2, 3*}$ \quad Thomas Edward Yerxa$^{2, 4}$ \\ \quad \textbf{Miah N. Pitcher}$^{5}$ \quad \textbf{Marla Feller}$^{5, 6}$  \quad \textbf{SueYeon Chung}$^{2,4}$\\
        $^1$Biophysics Graduate Group, University of California, Berkeley,\\
        $^2$Center for Computational Neuroscience, Flatiron Institute,\\
        $^3$Courant Inst. of Mathematical Sciences, New York University,\\
        $^4$Center for Neural Science, New York University,\\
        $^5$Helen Wills Neuroscience Institute, University of California, Berkeley,\\
        $^6$Department of Molecular and Cell Biology, University of California, Berkeley\\
        $^*$Equal contribution \\
        \texttt{ligeralde@berkeley.edu}, \texttt{yilun.kuang@nyu.edu}\\
}

\begin{document}

\maketitle

\begin{abstract}
  Prior to the onset of vision, neurons in the developing mammalian retina spontaneously fire in correlated activity patterns known as retinal waves. Experimental evidence suggests that retinal waves strongly influence the emergence of sensory representations before visual experience. We aim to model this early stage of functional development by using movies of neurally active developing retinas as pre-training data for neural networks. Specifically, we pre-train a ResNet-18 with an unsupervised contrastive learning objective (SimCLR) on both simulated and experimentally-obtained movies of retinal waves, then evaluate its performance on image classification tasks. We find that pre-training on retinal waves significantly improves performance on tasks that test object invariance to spatial translation, while slightly improving performance on more complex tasks like image classification. Notably, these performance boosts are realized on held-out natural images even though the pre-training procedure does not include any natural image data. We then propose a geometrical explanation for the increase in network performance, namely that the spatiotemporal characteristics of retinal waves facilitate the formation of separable feature representations. In particular, we demonstrate that networks pre-trained on retinal waves are more effective at separating image manifolds than randomly initialized networks, especially for manifolds defined by sets of spatial translations. These findings indicate that the broad spatiotemporal properties of retinal waves prepare networks for higher order feature extraction.
\end{abstract}

\section{Introduction}
The visual system has an extraordinary capacity for rapidly and accurately recognizing distinct objects in the face of identity-preserving transformations \citep{Thorpe1996, dicarlo_how_2012, Rajalingham2015}. Neural recordings suggest this is in part possible due to the high degree of linear separability between neural responses to different stimuli \citep{Hung2005, dicarlo_how_2012}. Interestingly, deep convolutional neural networks (DCNNs) trained to classify images can perform invariant object categorization at near human-level accuracy \citep{Cadieu2014} and have been shown to exhibit representations similar to neural activities in mammalian systems \citep{Yamins2014, wen2018deep}. Furthermore, DCNN layers have analagous properties to the visual hierarchy, whereby feature transformations at each layer induce linear separability in the object manifolds \citep{cohen_separability_2020}. DCNNs therefore offer a useful testbed for modeling the visual system \citep{KhalighRazavi2014, Kheradpisheh2016, Yamins2016, Wen2018}. However, the supervised learning methods used to train these models are unlikely to explain how the brain learns object recognition, given that large amounts of labeled examples are not necessary for visual development \citep{Bergelson2012, Bergelson2017, frank2021variability, zhuang_et_al_2021}. In this work, we explore the potential of innate neural activity as pre-training data for DCNNs and ask whether the internal representations that enable object recognition can be learned without access to any external visual information. 

The motivation for this work is grounded in developmental neurobiology. Many key aspects of visual system organization are well-established before visual experience, such as topographic maps, orientation selectivity, and ocular dominance \citep{espinosa_development_2012}. Notably, axon targeting can largely be learned by innately generated signals such as spontaneous neural activity and molecular guidance cues \citep{feller_chapter_2020}. These findings suggest external stimuli are unnecessary for the initial development of the early visual system. 

Here, we investigate whether a particular form of spontaneous activity known as retinal waves can instruct formation of the feed-forward connections that support object recognition. Retinal waves are a developmental phenomenon characterized by correlated patterns of propagating, network-level activity among groups of retinal ganglion cells (RGCs) prior to eye-opening \citep{arroyo_spatiotemporal_2016}. Experimental and computational evidence suggests that retinal waves instruct the formation of retinotopic maps, enabling RGC axons to reach their targets in the superior colliculus and lateral geniculate nucleus before the onset of visual experience \citep{cang_development_2005, chandrasekaran_evidence_2005, huberman_spontaneous_2006, markowitz_retinal_2012, hunt_sparse_2012, choi_building_2021}. 

Our core finding is that DCNNs pre-trained on movies of retinal waves produce more linearly separable representations of natural images compared to randomly initialized networks. To demonstrate this, we pre-train the hidden layers in a ResNet-18 classifier on calcium imaging movies of whole developing mouse retinas using the SimCLR \citep{pmlr-v119-chen20j} objective. This task-independent phase is meant to simulate the experience-independent period of visual development prior to eye-opening. We then evaluate network performance on a set of image classification tasks and find that networks pre-trained on retinal wave timecourses consistently outperform random controls. We explain this performance increase using the framework of manifold geometry \citep{chung_neural_2021}. Specifically, we characterize the geometry of the networks' internal feature representations \citep{chung_classification_2018} and find that networks pre-trained on retinal waves more effectively separate object manifolds. We also find the extent of separability is task-dependent and most pronounced for tasks that test spatial invariances. Our results suggest that the spatiotemporal information in retinal waves is relevant for object recognition in natural scenes and point towards an instructive role for retinal waves during early synapse formation in visual circuits.

\section{Methods}
\begin{figure}
\centering
\includegraphics{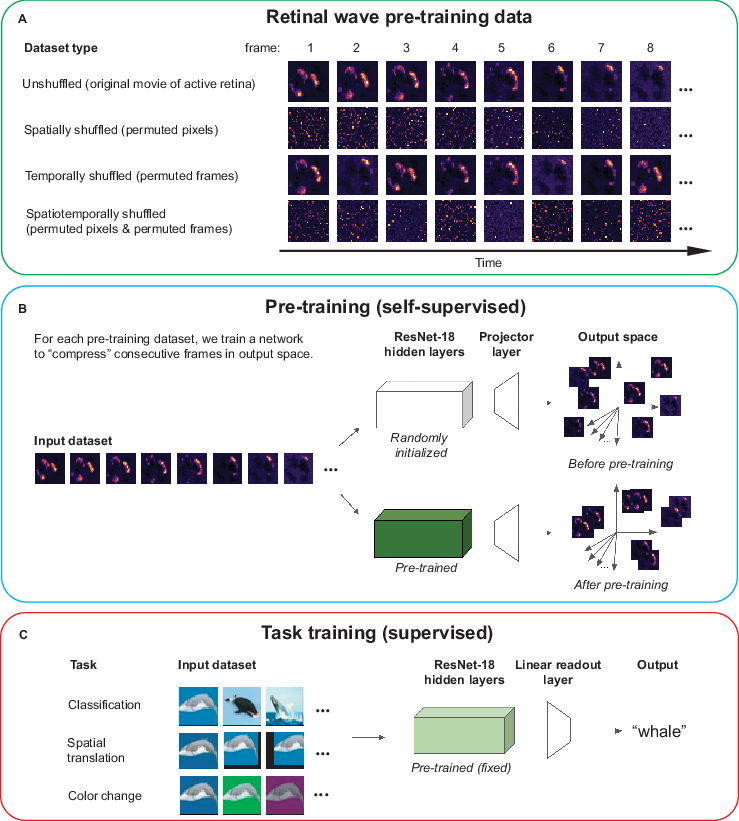}
\caption{\textbf{Network training pipeline.} (A) Retinal wave movies and three permutations of the original movies are used as pre-training datasets (B) contrastive learning to learn temporally close spatial correlations (C) evaluation of network performance on three labeling tasks.}
\label{fig:training_schematic}
\end{figure}
\subsection{Pre-training}
To test whether spatiotemporal features of retinal waves learned during pre-training will improve performance on visual tasks, we follow the pipeline described in Fig. \ref{fig:training_schematic}. Given a movie of a neurally active developing retina (Fig. \ref{fig:training_schematic}A), we first train a ResNet-18 to compress temporally consecutive frames of the movie in output space, while pushing apart temporally distant frames (Fig. \ref{fig:training_schematic}B) using the SimCLR training objective \citep{pmlr-v119-chen20j}. This is in accordance with the finding that temporally close activity bursts convey the most spatial information about relative RGC position \citep{Butts2001, Butts2002}. This phase is meant to simulate the period of cortical development \textit{prior} to visual experience. We pre-train two kinds of networks: the first using macroscope movies of retinal waves obtained via calcium imaging of whole retinas dissected from postnatal mice (for experimental methods, see Supplementary Material S0), and the second using simulated movies of retinal waves from a parametrized, reaction-diffusion based model \citep{Lansdell2014}. 
To isolate the effects of the spatial and temporal characteristics of retinal waves, we pre-train networks on three additional types of datasets created by modifying the original movies: \textbf{spatially shuffled}, in which the pixels of each frame are randomly permuted; \textbf{temporally shuffled}, in which the frame order is randomly permuted; and \textbf{spatiotemporally shuffled}, in which both the pixels of each frame and the frame order are randomly permuted (Fig. \ref{fig:training_schematic}A). Spatially shuffled waves contain information about how the overall distribution of RGC activities changes over time, but lack the continuously varying spatial structure present in the original movies. As such, spatially shuffled pre-training controls for how much task information can be inferred only through temporally local changes in the population statistics of RGC activity. Comparing temporally shuffled and spatiotemporally shuffled waves controls for the amount of task-relevant, temporally non-local information in retinal waves. If correlations between temporally distant frames are relevant for a given task, networks trained on temporally shuffled waves should perform better than those trained on spatiotemporally shuffled waves. We compare all pre-training conditions to a He random initialized control network that has not been pre-trained, for a total of nine conditions. 

\textbf{Preprocessing: }To filter out calcium transients, periods of inactivity, and random noise in the calcium imaging data, watershed image segmentation is used to identify periods of continuous retinal wave activity spanning a given number of frames, with each period denoted as a \say{wave event}. We use publicly available code for watershed segmentation from \url{https://github.com/Llamero/Feller_Retinal_Wave_Analysis}. We aggregate movies from four retinas, resulting in $\sim$60,000 total frames of real retinal wave pre-training data. Simulated retinal wave data is generated using the model in \cite{Lansdell2014} \say{out-of-the-box}. The area parameter of the simulation is changed to match the area of the isolated real retinas, and the \say{strength} parameter $\alpha$ is modified to 0.5 to increase the wave frequency and eliminate long periods of inactivity. The model frame rate is matched to that of the macroscope data. The model is run to obtain a total of $\sim$237,000 frames of simulation data. Because the simulated data is far less noisy than the real data, wave events are simply taken as the sets of frames in between periods of cell inactivity, without the need for image segmentation.

\textbf{Hyperparameters: }Networks are pre-trained with a projector layer \citep{zbontar2021barlow} of dimensions $8192\times8192\times8192$ for 100 epochs with a learning rate of 0.0001 and Adam optimization based on a grid hyperparameter search. 
% Following , the projector consists of two linear layers of output dimension 8192 followed by batch normalization and a ReLU non-linearity. A third linear layer of output dimension 8192 is added on top of the two linear layer blocks.  
Because wave events occur for varying lengths of time, batches are formed by randomly sampling whole wave events from the movie until the total number of sampled frames exceeds a threshold value of 3000. Positive examples are defined as consecutive frames within the same wave event, and negative examples are defined as all frames outside of that wave event. 

\subsection{Task training}
To test the effects of pre-training on task performance, we add a linear readout layer to the pre-trained weights and train linear readout layer weights on labeled images while leaving the pre-trained hidden layer weights fixed (Fig. \ref{fig:training_schematic}C). This phase is meant to simulate a test of the functionality gained from retinal wave activity at the onset of visual experience. We use this procedure to evaluate network performance on three labeling tasks. We report the mean and standard deviations for test accuracy across the three seeded random network initializations. 

\textbf{Classification task: }The first task is standard image classification on CIFAR-10. 

\textbf{Spatial translation task: } For the second task, we train networks to classify spatially translated images drawn from CIFAR-100. To generate the task data, we first choose 10 of 100 classes at random and draw a random image from each class, which we denote as a \say{base} image. An image in the task dataset is then generated as a random affine transformation (up to 16 pixels in the $x$ and $y$ directions) of one of the 10 base images. Using this procedure, each base image is used to generate 5000 training images and 1000 test images, for a total of 50,000 training images and 10,000 test images. The networks are trained to classify a given training image with the label of its original base image. 

\textbf{Color change task: }For the third task, we train networks to classify recolorations of the same 10 base images used in the spatial translation task. The task data is generated by the same procedure, only instead of random affine transformations, we apply random color transformations to the base image that range from 50 to 100\% changes in saturation, brightness, contrast, and hue. The networks are trained to classify a given training image with the label of its original base image. 

\textbf{Hyperparameters: }In task training, the projector dimension used in pre-training is removed and replaced with a $512 \times 10$ linear readout layer \citep{zbontar2021barlow}. The readout layer is trained for 100 epochs, batches of size 100, and learning rate of 0.0001 on 50,000 labeled training images. The performance is evaluated on 10,000 labeled test images. 

\subsection{Manifold analysis}
\label{section:manifold_methods}
\begin{figure}[!htbp]
\centering
\includegraphics{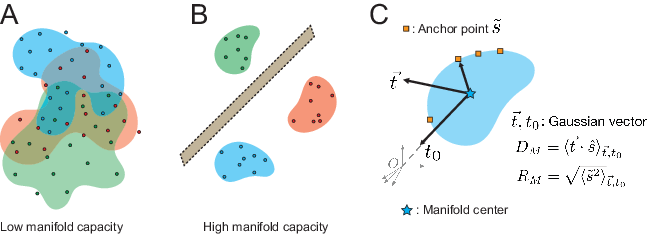}
\caption{\textbf{Illustration of point cloud manifolds}. (A) Tangled manifolds exhibit low capacity (B) untangled manifolds exhibit high capacity and are separable by a hyperplane (C) manifold dimension measures spread of anchor points across the manifold axes by projection of a Gaussian vector onto an anchor point. Manifold radius measures the norm of an anchor point in the manifold subspace.}
\label{fig:manifold_schematic}
\end{figure}
The set of neural responses to different presentations of a given stimulus define a neural object manifold. The linear separability of these manifolds as a function of their geometry enables discrimination between stimuli \citep{chung_neural_2021}. We use this framework to characterize how pre-training on retinal waves changes the geometry of representation. We examine three quantities of manifolds that determine their separability, namely the capacity $\alpha_c$, the dimension $D_M$, and the radius $R_M$.

\textbf{Capacity $\alpha_c$: }We consider a set of $P$ object manifolds linearly separable if they can be classified into binary classes by a hyperplane in $N$-dimensional feature space. The theory of manifold geometry shows that the value of the manifold capacity $\alpha_c$ determines the extent of separability in the limit of large $P$ and $N$: if $P/N < \alpha_c$, the manifolds are separable with high probability; if $P/N > \alpha_c$, the manifolds are inseparable with high probability. Therefore, the higher the value of $\alpha_c$, the higher the probability of separability for a given set of manifolds (Figs. \ref{fig:manifold_schematic}A,B). For point-cloud manifolds, in which each manifold consists of $M$ data points each corresponding to an example of the given object, the capacity can be shown to be bounded as $\frac{2}{M} \leq \alpha_c \leq 2$ \citep{chung_classification_2018}. The theory of manifold geometry also shows that capacity is determined by two quantities which describe the geometry of the object manifolds in $N$-space: the dimension $D_M$ and the radius $R_M$. These are statistical quantities defined for each manifold by considering the spread of points in the manifold's convex hull, called anchor points, over variations in the manifold's labeling and location in $N$-space (Fig. \ref{fig:manifold_schematic}C). For large $N$, $\alpha_c$ is inversely proportional to $\sqrt{D_M}$ and $R_M$ \cite{chung_statistical_2021}. All three quantities --- $\alpha_c$, $D_M$, and $R_M$ --- are estimated using algorithms based on statistical mechanical mean-field techniques described in \citep{Stephenson2020UntanglingII}. 

\textbf{Dimension $D_M$: } Dimension is the spread of anchor points across the manifold axes and estimates the average embedding dimension of the manifold (Fig. \ref{fig:manifold_schematic}C). 

\textbf{Radius $R_M$: }Radius is the average distance between the manifold center and anchor points and reflects the scale of the manifold compared to the overall data distribution. (Fig. \ref{fig:manifold_schematic}C). 

\textbf{Simulation capacity $\alpha_{sim}$: } We note that $\alpha_c$ is a theoretical estimate of linear separability that may deviate from the true capacity in the regime of finite manifolds $P$ and feature dimensions $N$ \citep{chung_classification_2018}. Simulation capacity provides a numerical approximation of the ground-truth manifold capacity. We calculate simulation capacity by first running linear classifications with fixed $P$ and varying $N$ until the probability of manifold separation converges to 0.5. The final value of $N = N_c$ is used to calculate the simulation capacity $\alpha_{sim}=P/N_{c}$. We report the correspondence between $\alpha_c$ and simulation capacity in Fig. S2. 

\textbf{Task data manifolds: }To examine how pre-training with retinal waves affects the geometry, and in turn the separability, of neural object manifolds for each task, we extract the network activations at each ReLU layer for $P = 50$ manifolds consisting of $M = 20$ examples. For \textbf{standard classification}, each manifold corresponds to an image class in CIFAR-100. Examples for each manifold are drawn from the given class based on the ranked 20-highest softmax probability scores output by a well-trained classifier. For both \textbf{spatial translation} and \textbf{color change}, each manifold corresponds to one random base image drawn from CIFAR-100. Examples for each spatial translation manifold are generated by applying random affine shifts up to 3 pixels in both directions to the base image. Examples for each color change manifold are generated by applying random $50-150\%$ changes in saturation, brightness, hue, and contrast to the base image. We also measure the capacity and geometry of the manifolds defined by retinal waves, where each manifold consists of frames belonging to a given wave event (Fig. S1).

We report the mean and standard deviations for all manifold quantities across three seeded random network initializations. For all manifold analysis, we use publicly available code from \url{https://github.com/schung039/neural_manifolds_replicaMFT}. 

Pre-training, task training, and manifold analysis was done on an internal cluster using NVIDIA 16 GB V100 (Volta) GPUs. All code for pre-processing, pre-training, task training, and analysis is available at \url{https://github.com/chung-neuroai-lab/retinal_waves_learning}.

\section{Results}
\begin{figure}
\centering
\includegraphics{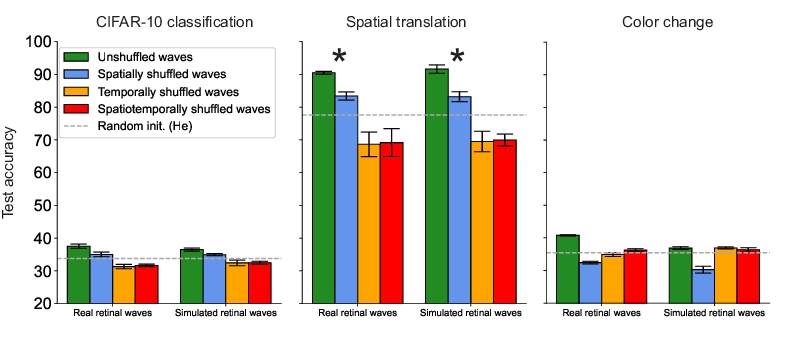}
\caption{\textbf{Test accuracy for pre-trained networks in three labeling tasks.} Asterisks indicate that the performance increase from pre-training on retinal waves is highest for the spatial translation task.}
\label{fig:classifier}
\end{figure}
\subsection{Pre-training on retinal waves improves task performance}
\label{section:accuracy}
Our main result is that self-supervised pre-training of networks on movies of retinal waves improves object separability for labeled natural images. We find that pre-training on the original, unshuffled wave movies yields the highest performance increase in the spatial translation task (Fig. \ref{fig:classifier}). This suggests that retinal waves contain information that supports learning object invariance to spatial translation. Pre-training on spatially shuffled waves yields a moderate improvement above random initialization in this task, suggesting that learning temporally local changes in the overall distribution of activities is also relevant for this function. Destroying the temporal structure of the waves, however, yields performance below random initialization, as shown in the temporally and spatiotemporally shuffled pre-training conditions. This suggests that temporally local, rather than global correlations in retinal waves are most relevant for learning spatial invariance. This is consistent with the previous finding that little information is gained by considering RGC activity bursts more than 3 sec (around 35 frames) apart \citep{Butts2001, Butts2002}. These networks perhaps even learn non-local features that actually hinder task learning, as suggested by their below-random-network performance. We further explore this idea in Sec. \ref{section:correlation}.

Classification is a far more complex task than spatial translation as it requires mapping visual information onto higher level semantic structures, information not present retinal waves. Accordingly, performance for this task is significantly lower for pre-trained networks overall than for spatial translation. However, networks trained on unshuffled waves still perform slightly better than the others (Fig. \ref{fig:classifier}). A similar trend emerges for the color change task, for which we also did not expect pre-training to yield any advantage. A potential reason for the performance increases in both cases is the persistence of similar features across examples in the same class. Visual patterns like edges and curves are features that retinal waves may train the visual system to recognize \citep{albert_innate_2008}. We further explore reasons for these small performance boosts in \ref{section:correlation}.

While accuracy provides a proxy for the task-specific relevance of retinal waves, it does not give insight into how retinal waves influence learned feature representations. In the next section, we address this question by examining the geometry of task object manifolds across pre-training conditions. 

\subsection{Pre-training on retinal waves increases separability for manifolds defined by invariance to spatial translation}
\begin{figure}
\centering
\includegraphics{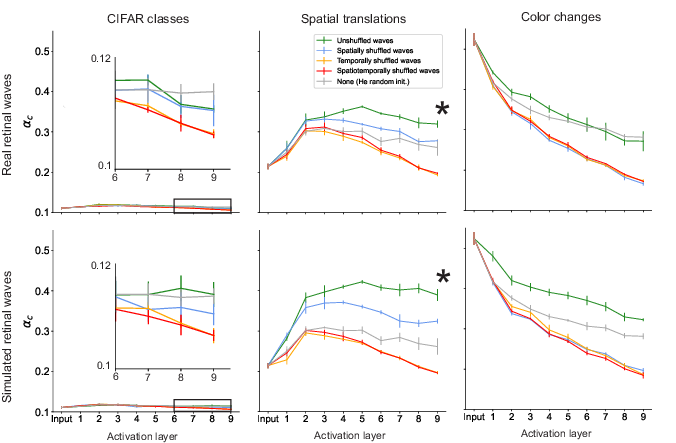}
\caption{\textbf{Changes in classification capacity over network layers.} Asterisks indicate that the capacity of spatial translation manifolds increases the most along the hierarchy of the network pre-trained on unshuffled retinal waves. }
\label{fig:capacity}
\end{figure}

\begin{figure}[h]
\centering
\includegraphics{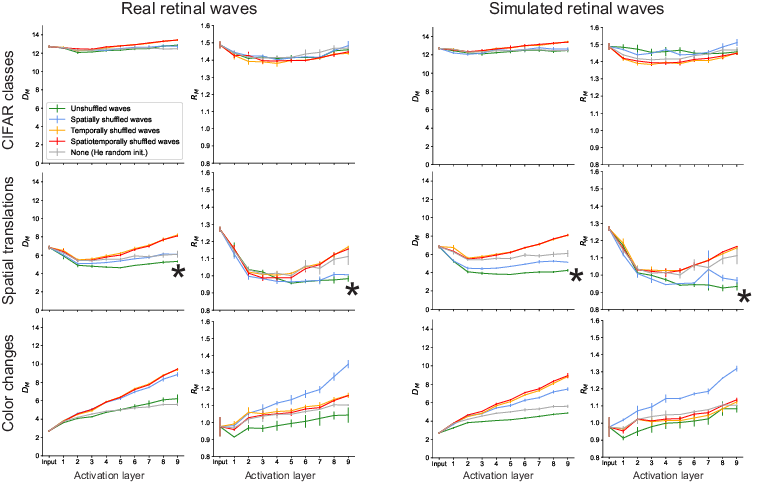}
\caption{\textbf{Changes in manifold geometry over network layers.} Asterisks indicate that networks pre-trained on unshuffled retinal waves most effectively compress spatial translation manifolds.}
\label{fig:dim-rad}
\end{figure}

Previous work shows that DCNNs trained to classify images increase the object manifold capacity from the input to output layers \citep{cohen_separability_2020}. We only observe this behavior for the spatial translation manifold. Consistent with the accuracy results, networks trained on unshuffled waves and spatially shuffled waves yield increases in capacity relative to randomly initialized networks, while networks trained on temporally and spatiotemporally shuffled waves do not substantially change the capacity between the input and output layers (Fig. \ref{fig:capacity}). As expected, the spatial translation manifolds in networks pre-trained on unshuffled waves also have lower dimension and radius compared to those in the other networks, while networks pre-trained on spatially shuffled waves only appear to decrease the radius (Fig. \ref{fig:dim-rad}). These results suggest that pre-training on retinal waves has a direct influence on the geometry and separability of neural object manifolds for tasks that involve learning spatial invariance.

In all networks, the capacity of the CIFAR class manifold (see inset, Fig. \ref{fig:capacity}) remains nearly constant around the theoretical lower bound of 0.1 (Sec. \ref{section:manifold_methods}).  All networks also yield a decrease in capacity for the color change manifold at each successive layer (Fig. \ref{fig:capacity}). (Although the network trained on simulated unshuffled waves appears to have a relatively high capacity for the color change manifold, this particular value actually overestimates the ground truth simulation capacity, which we show in Fig. S2). The dimensions and radii of the CIFAR class and color change manifolds also do not show any consistent ordering that points to a clear advantage of pre-training on retinal waves relative to the random baseline (Fig. \ref{fig:dim-rad}). These results are consistent with the poor accuracy in these tasks across all networks. However, if pre-training does not substantially affect these object manifolds, what accounts for the slight boost in performance on these tasks for the networks pre-trained on unshuffled waves? To address this question, we explore two factors external to the geometry of individual manifolds, namely the inter-manifold correlation and the effective dimensionality of the feature space. 
\subsection{Pre-training on retinal waves decreases inter-manifold correlations and maintain effective dimensionality}
\label{section:correlation}
\begin{figure}
\centering
\includegraphics{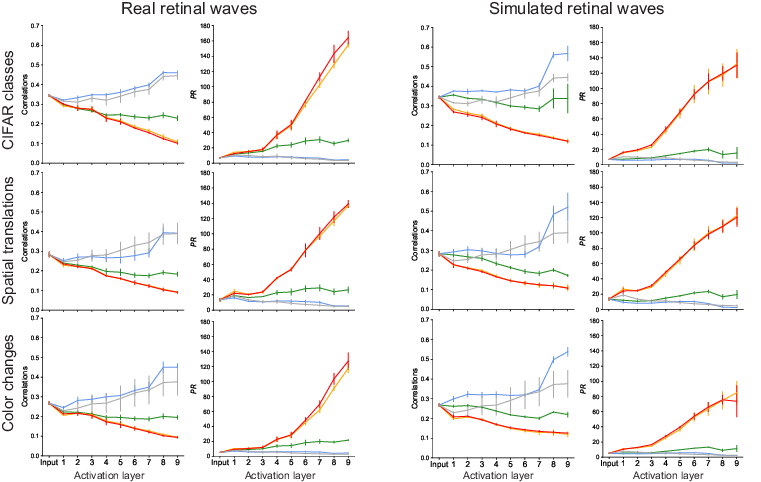}
\caption{\textbf{Changes in inter-manifold correlation and participation ratio along network layers}. Only the network pre-trained on unshuffled waves consistently reduces correlation and avoids vanishing/exploding dimensionality.}
\label{fig:corr-pr}
\end{figure}

A high degree of correlation between manifold centers may lead to clustering of object manifolds in feature space, making them more difficult to separate and decreasing the effective capacity. Previous work demonstrates that training DCNNs leads to decorrelation of the manifold centers \citep{cohen_separability_2020}. Here, we measure the pairwise correlation coefficient between manifold centers at each network layer and find that networks pre-trained on unshuffled retinal waves decrease center correlations relative to randomly initialized networks and networks pre-trained on spatially shuffled waves for all three tasks (Fig. \ref{fig:corr-pr}). Unshuffled pre-training also leads to a generally consistent decrease in correlation along at each successive network layer. Interestingly, temporally and spatiotemporally shuffled pre-training also produce networks that exhibit this behavior, in addition to having lower correlations than in the unshuffled case. However, based on their poor task performance and low capacities of their feature representations, it is likely this is simply due to the explosion in dimensionality of their respective feature spaces, which we discuss next.

Ideally, a well-trained classifier will extract the features that correspond to the highest sources of variance in the data, while separating out low-variance features that do not correspond to meaningful distinctions between samples. Participation ratio ($PR$) varies from 1 to $N$ and measures how data variance is spread out across the feature dimensions: if $PR = 1$, the variance is concentrated entirely in one feature; if $PR = N$, the variance is spread out evenly across all features \citep{Gao2017}. In general, a good classifier will maintain a $PR > 1$ in the feature dimensions so as not to destroy the structure in the data, while also keeping $PR < N$ so as to preserve only the meaningful features that capture the latent dimensionality of the data. The layer-wise participation ratio suggests that networks pre-trained on unshuffled waves maintain this happy medium in all three tasks (Fig. \ref{fig:corr-pr}). Networks pre-trained on spatially shuffled waves decrease participation ratio to near the lower bound, consistent with the idea that they broadly capture population-level statistics, but fail to learn many spatially local features that likely lie along other dimensions. The large increase in $PR$ observed in networks trained on temporally and spatiotemporally shuffled waves suggests that they do in fact learn features that are not relevant for the task dataset, as proposed in Section \ref{section:accuracy}. These extraneous features would account for the increase in $PR$ above the values observed in other networks. Notably, correlation and $PR$ are inversely related, suggesting that high effective dimensionality is a factor in separation of manifold centers. The trends observed in $PR$ are consistent with the trends in layer-wise explained variance, which measures how many feature dimensions account for a given percentage of variance in the data (Fig. S3).

\section{Discussion}
To our knowledge, this is the first computational work that directly explores how real retinal waves can influence neural object representations, demonstrating a bioplausible means of learning spatial invariance without training on large datasets of labeled images. While DCNNs trained on labeled images achieve state-of-the-art performance and even predict neural responses \citep{Yamins2014, Schrimpf2018}, these models are unlikely to explain how biological vision develops. Unsupervised and self-supervised learning mechanisms have therefore been proposed as biologically plausible means of learning object recognition \citep{zhuang_et_al_2021}. However, standard implementations of these algorithms still require natural images or videos as training inputs, which effectively simulate a visual experience. Though visual experience certainly shapes cortical functional development \citep{Pecka2014, Matteucci2020, Kowalewski2021, Nishio2021}, models that wholly rely on image data do not account for the functionality, connectivity, and feature selectivity already observed in animals prior to the onset of vision \citep{Sherk1976, Ko2014, ackman_role_2014, Xu2016, tiriac_retinal_2021}. Consistent with our results, previous work has demonstrated that self-supervised learning on structured noise can improve classification accuracy on unseen images \citep{raghavan_neural_2019, raghavan_self_org, baradad2021learning}. Additionally, simulated retinal waves have been shown to yield V1-like receptive fields when used as inputs for sparse coding algorithms \citep{albert_innate_2008, hunt_sparse_2012, behpour_role_2021} and slow feature analysis \citep{dahne_slow_2014}. 

We demonstrate that pre-training on retinal waves has two primary effects on learned representations that can account for increases in task performance. The first is an increase in the separability of individual object manifolds. This effect is pronounced in the spatial translation task, suggesting that the spatiotemporal characteristics of retinal waves train networks to learn spatial translation invariance. To show this, we analyze the geometry of the neural object manifolds defined by affine transformations of a single object (image) and find they are more linearly separable when represented in networks pre-trained on unshuffled retinal waves (Figs. \ref{fig:capacity}, \ref{fig:dim-rad}). Both the spatial and temporal characteristics of retinal waves are necessary for learning this task, as pre-training on spatially and/or temporally shuffled retinal waves leads to poor separability of spatial translation manifolds. Pre-training does not have a significant effect on the separability of the manifolds defined by CIFAR image classes or color changes of a single object (Figs. \ref{fig:capacity}, \ref{fig:dim-rad}), suggesting a qualitative bound on the scope of tasks for which retinal waves are useful training signals. 

We also observe that pre-training on retinal waves reduces center correlations between neural object manifolds and increases the effective dimensionality of the feature space (Figs. \ref{fig:corr-pr}). Both effects are directly correlated with linear separability and appear to be independent of the effect on individual manifold separability, as they are observed in all three tasks. 

Together, these two effects of pre-training on retinal waves correspond to distinct \textit{local} and \textit{global} mechanisms of transforming object representations, both of which are important for separability. At the local level, pre-training increases the compressibility of individual neural object manifolds, as shown in the increase in capacity and the concurrent decreases in dimension and radius. At the global level, pre-training places neural object manifolds in higher dimensional feature space, as shown by the increase in participation ratio and concurrent decrease in center correlation. These two regimes point to distinct ways in which retinal waves may influence emerging sensory representations.

We do not observe a significant difference between pre-training on real versus simulated retinal waves from the model. The advantage of the model is that we can generate an arbitrarily large set of pre-training data, at the risk of introducing free parameters that may lead to deviations from real data. Though we do not perform a direct comparison between the simulated and real data in this work, no clear difference emerges between these two datasets in terms of model performance or the geometry of the object representations. This suggests that for the tasks considered, the common features of these datasets --- such as spatiotemporal continuity between frames --- are the primary drivers of the observed effects. In future work, the model may be a useful tool for examining the effect of changing the waves' spatiotemporal characteristics on representation learning.

We note that our findings are subject to our choice of network architecture (ResNet-18), learning algorithm (SimCLR), and dataset (postnatal mouse retinal waves). Retinal waves occur during multiple stages of development \citep{Voufo2023} and drive formation of visual circuitry in numerous ways \citep{arroyo_spatiotemporal_2016}. Retinal waves are also not the only form of spontaneous activity during development \citep{Hanganu2006}. Along this line of work, future studies may consider the role of cortical feedback \citep{Murata2016}, introduce bioplausible, synaptically local learning rules \citep{illing2021local}, or investigate the role of spontaneous activity in other modalities like temporal prediction \citep{Luczak2022}. Additionally, laboratory experiments that test object recognition in mice \citep{Zoccolan2009} performed at the onset of vision could verify our model predictions and provide richer insight into the capacity of neural object manifolds during this early developmental period.

\begin{ack}
This work was funded by the Center for Computational Neuroscience at the Flatiron Institute of the Simons Foundation. AL was supported by the Flatiron Machine Learning X Science Summer School and the National Science Foundation under grant DGE 1752814. YK was supported by the NYU Center for Data Science Fellowship. MNP and MF were supported by the National Institutes of Health under grands NIH RO1EY013528, RO1EY019498, and P30EY003176. TY has no competing interests or relevant funding to declare. SC was supported by the Klingenstein-Simons Award. All experiments were performed on the Flatiron Institute high-performance computing cluster. 

%todo: ack
%todo: fix citation 36
\end{ack}
%%%%%%%%%%%%%%%%%%%%%%%%%%%%%%%%%%%%%%%%%%%%%%%%%%%%%%%%%%%%
\bibliography{unireps_2023}
\newpage

\setcounter{section}{-1}
\renewcommand{\thesection}{S\arabic{section}}

\setcounter{figure}{0}
\renewcommand{\figurename}{Figure}
\renewcommand{\thefigure}{S\arabic{figure}}

\section{Retina preparation and epifluorescent macroscope calcium imaging of retinal waves}
We obtain real retinal wave movies using the following procedure. Mice aged postnatal day 8 to 11 (P8 to P11) are deeply anesthetized with isoflurane inhalation and euthanized by decapitation. Eyes are immediately enucleated and retinas are dissected at room temperature under infra-red illumination in oxygenated (95\% \ce{O2} / 5\% \ce{CO2}) ACSF (in mM, 119 NaCl, 2.5 KCl, 1.3 \ce{MgCl2}, 1 \ce{K2HPO4}, 26.2 \ce{NaHCO3}, 11 D-glucose, and 2.5 \ce{CaCl2}). Cuts along the chloride fissure are made prior to isolating the retina from the retinal pigmented epithelium. These cuts are made to precisely orient the retina. The isolated retina is mounted whole on filter paper with the photoreceptor layer side down, and transferred in a recording chamber of a microscope for subsequent imaging. The whole-mount retina is continuously perfused (3 mL/min) with oxygenated ACSF media at 32 to 34°C for the duration of the experiment. Epifluorescent whole-retina calcium imaging is obtained using transgenic mice that express GCaMP6s in all retinal ganglion cells (Vglut2::GCaMP6s). Waves are imaged on a custom built macroscope (4$\times$ 0.28 NA objective, a 498Hz kinetix camera, 4.7 mm $\times$ 4.7 mm FOV, and 1.5 µm/pixel) controlled by µManager 2.0 software. The oriented retina is mounted onto nitrocellulose filter paper (Millipore) for a darker background. Waves are imaged for 30 minutes at a 12.5 Hz frequency and pixels are binned 4 $\times$ 4 for a resolution of 5.9 µm/pixel. GCaMP6s excitation is evoked with a 476 nm LED. 

\section{Analysis of retinal wave manifolds}
\begin{figure}
\centering
\includegraphics[scale=0.68]{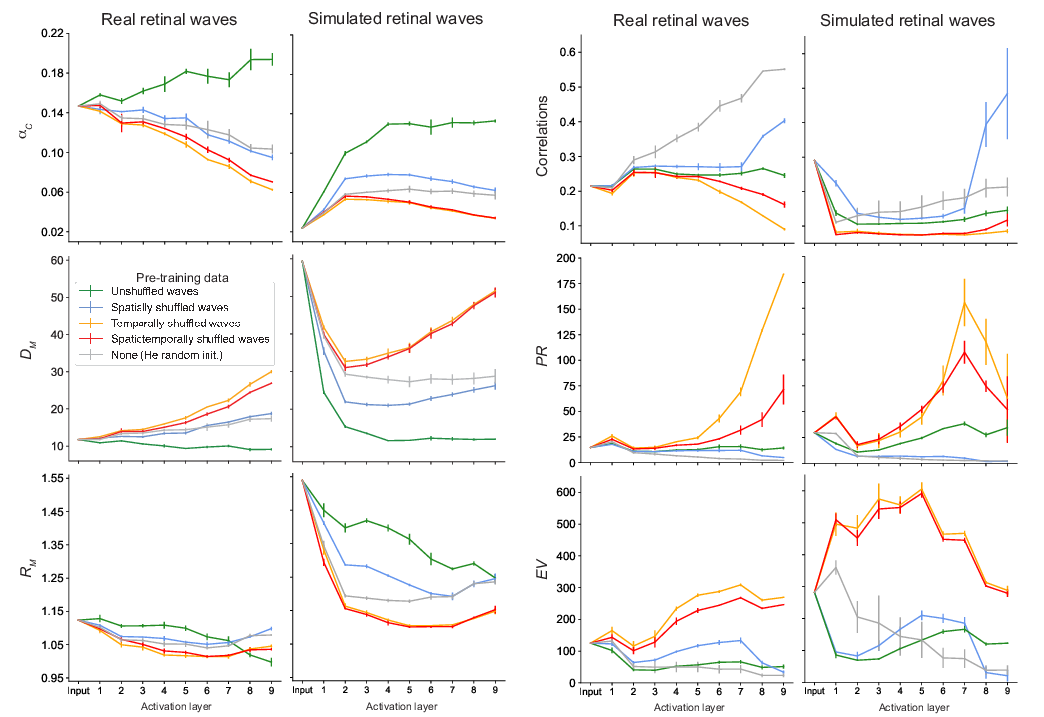}
\caption{\textbf{Changes in (unshuffled) wave manifolds over network layers.}}
\label{fig:wave_manifold}
\end{figure}

A wave manifold as described in Fig. \ref{fig:wave_manifold} is defined from a set of 50 frames from a randomly chosen single wave event in the original, unshuffled wave movie. We consider 50 such manifolds for all such metrics computed in Fig. \ref{fig:wave_manifold} across the five pre-training conditions. Explained variance is the number of dimensions in feature space that account for 90\% of the variance in the frames considered for manifold analysis.
As expected, networks pre-trained on unshuffled waves yield the highest capacity amongst all pre-training conditions for the wave manifolds. Interestingly, the manifolds for real retinal waves appear to have higher capacity at all layers compared to those for simulated waves. This may be due to the smaller size of the real retinal waves dataset, which could lead to less variability across frames than in the simulated dataset. This explanation is consistent with the fact that the $PR$ and $EV$ for simulated waves is higher than for real waves. The trend in correlation, $PR$, and $EV$ for both real and simulated wave manifolds reflects that observed in the task manifolds (Fig. 6), whereby the networks pre-trained on unshuffled waves maintain higher feature dimensionality and lower center correlation than the random networks, without producing a dimensionality explosion like the networks pre-trained on temporally shuffled waves.

\section{Simulation capacity}
\label{section:simulation}
\begin{figure}
\centering
\includegraphics[scale=0.787]{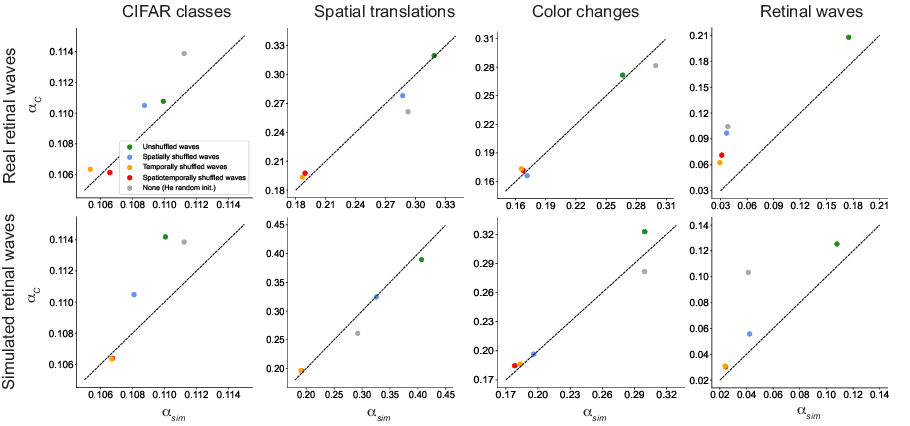}
\caption{\textbf{Correspondence between theoretical and simulation capacity.} Each point represents mean over three random network initializations at the last activation layer, with the exception of the retinal wave manifolds (last column), where the measurements are taken in the projector layer. Dotted gray line denotes exact match between $\alpha_c$ and $\alpha_{sim}$.}
\label{fig:th_sim}
\end{figure}

We report the correspondence between real and simulation capacity in the last activation layer (Fig. \ref{fig:th_sim}). We observe a high degree of correspondence between these values with exception of the overestimation of $\alpha_c$ for the color change manifold in networks pre-trained on simulated retinal waves (second row, third column). We also note that for retinal wave manifolds, calculation of $\alpha_{sim}$ is numerically unstable in the last activation layer for networks not trained on unshuffled waves (fourth column). This may occur when the manifold capacity is low relative to the feature dimension $N$, resulting in poor separability. For this reason, we instead report the values of $\alpha_c$ and $\alpha_{sim}$ for the wave manifolds in the projector layer (see Sections \ref{section:architecture} for more details).

\section{Explained variance of task manifolds} 
\begin{figure}
\centering
\includegraphics{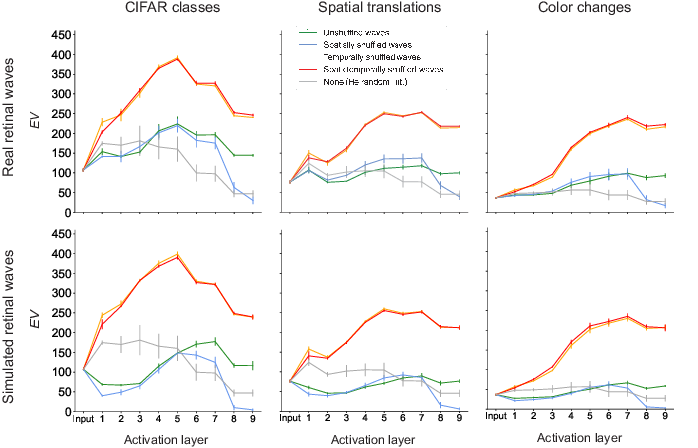}
\caption{\textbf{Changes in explained variance over network layers.} y-axis denotes the number of feature dimensions that account for 90\% of data variance.}
\label{fig:ev}
\end{figure}
We measure explained variance by the number of dimensions in feature space that account for 90\% of the variance in the examples considered for manifold analysis (Fig. \ref{fig:ev}). The trend in $EV$ in all tasks reflects that observed in center correlation and $PR$ (Fig. 6), whereby the networks pre-trained on unshuffled waves maintain higher feature dimensionality and lower center correlation than the random networks, without producing a dimensionality explosion like the networks pre-trained on temporally shuffled waves.

\section{Model architecture and activation extraction}
\label{section:architecture}
For all pre-training, we use a ResNet-18 network \citep{he2015deep} (without the fully connected classification layer) followed by a projector layer. The projector consists of three linear layers with 8192 output units. The first two linear layers in the projector are each followed by a batch normalization layer and ReLU activations. The ResNet-18 backbone without the classification layer is sometimes referred to in self-supervised learning as an \say{encoder}, and the outputs of the projector layer are referred to as \say{embeddings} \citep{zbontar2021barlow}. The SimCLR loss is computed on the embeddings during pre-training, and during task training, the projector is swapped out with a $512 \times 10$ linear readout layer. This procedure of swapping out the projector has been shown empirically to be beneficial in transfer learning, where there is a misalignment between the pre-training and training tasks \citep{balestriero2023cookbook}. 

For all theoretical manifold quantities ($\alpha_c, D_M, R_M$, correlation, $PR, EV$) the outputs of the intermediate ReLU activations in the encoder (for a total of 9 activation layers \citep{he2015deep}) are extracted to analyze the internal representations of the task or wave manifolds at each layer. Due to the high computational cost, we only calculate the simulation capacity at the last ReLU in the encoder (unless otherwise stated; see Section \ref{section:simulation} for details). 

\section{Examples of real and simulated wave data} 
\begin{figure}
\centering
\includegraphics{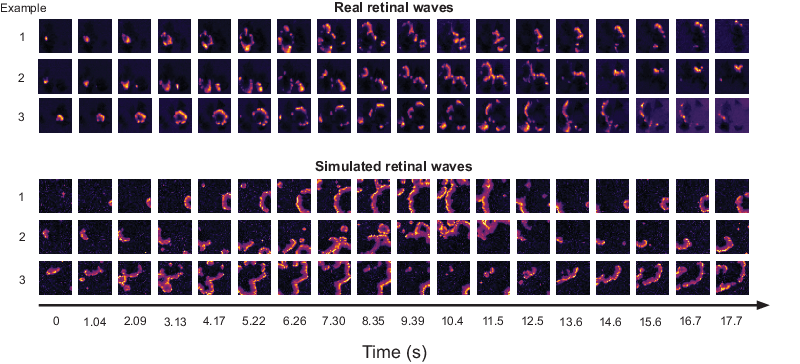}
\caption{\textbf{Qualitative comparison of representative examples from real and simulated retinal wave movies.} For each example, every 12th frame is presented in order to visualize wave activity over longer a period of time.}
\label{fig:wave_examples}
\end{figure}
\begin{figure}
\centering
\includegraphics[scale=0.8]{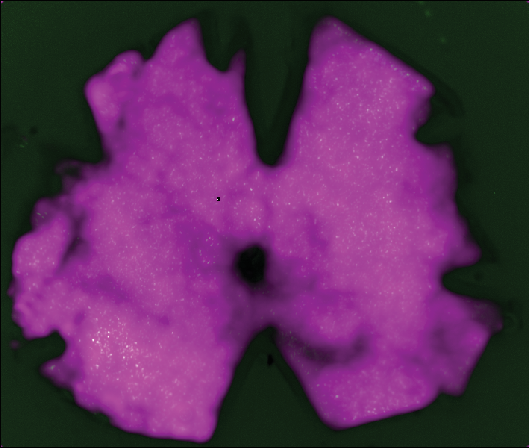}
\caption{\textbf{Area of an isolated retina used to obtain real retinal wave data.} Retina shown in pink. Pixel-wise area calculated using open source Fiji software and then converting to metric units based on macroscope resolution.}
\label{fig:retina}
\end{figure}
While we do not perform a direct quantitative comparison between the real and simulated retinal waves in this work, we present 3 representative examples from each dataset taken over a time period of about 18 sec. (Fig. \ref{fig:wave_examples}). A key difference between the two datasets is that in the real retinal wave movies, the waves must terminate when they reach a boundary of the imaged retina (Fig. \ref{fig:retina}), but in the simulated retinal wave movies, the model \say{retina} is a uniform surface that extends beyond the field of view \citep{Lansdell2014}. For this reason, in the simulated movies, the waves may continue past the frame. We partially adjust for this difference by setting the area parameter of the simulated retinal wave model as the average area of the calcium imaged retinas, though this adjustment does not account for any variations in wave characteristics induced by the retinal border. During pre-processing, both datasets are normalized to have a global mean and variance of 0 and 1, respectively. Both datasets are also given a new axis of size 3 and copied along this axis so that each color channel (RGB) of the network is pre-trained on the same data.  

\section{Spatial translation and color change base images}
\begin{figure}
\centering
\includegraphics{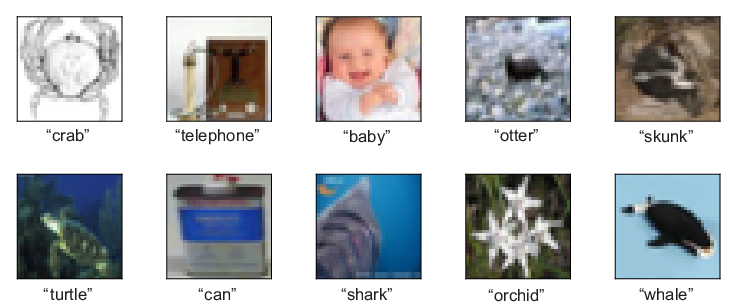}
\caption{\textbf{Base images and labels for spatial translation and color change tasks.}}
\label{fig:aug_task_exs}
\end{figure}
For both spatial translation and color change task training and testing, we use the same base images shown in Fig. \ref{fig:aug_task_exs}. These images are randomly sampled from CIFAR-100 \citep{krizhevsky2009learning}. Note that for generating the spatial translation and color change manifolds, we use a set of 50 randomly sampled base images that does not necessarily contain the base images used during task training and testing. 
\newpage
\end{document}